%% file: JFS_thrustaug.tex
\documentclass[preprint,authoryear,12pt]{elsarticle}

 \usepackage{graphicx}
 \usepackage{varioref}
 \usepackage{wrapfig}
 \usepackage{dcolumn}
  \newcolumntype{d}{D{.}{.}{-1}}
 \usepackage{subfigure}
 \usepackage{subfigmat}
 \usepackage{fancyvrb}
  \fvset{fontsize=\footnotesize,xleftmargin=2em}
 \usepackage{lettrine}
 \usepackage{amsmath}
 \usepackage{latexsym}
 \usepackage{color} 

\begin{document}

\begin{frontmatter}

\title{Thrust augmentation of flapping airfoils in low Reynolds number flow using a flexible membrane\tnoteref{t1}}
\tnotetext[t1]{An earlier version of this paper was presented as Paper 1640 at the 53$^{\rm rd}$ AIAA Structures, Structural Dynamics and Materials Conference (Honolulu, HI, April 2012).}
\author{Justin W. Jaworski\corref{cor1}}
\ead{jaworski@lehigh.edu}
\address{Department of Mechanical Engineering and Mechanics, Lehigh University, \\
Bethlehem, PA 18015-3085, USA}

\author{Raymond E. Gordnier}
\ead{raymond.gordnier@wpafb.af.mil}
\address{AFRL/RQAC, Bldg 146 Rm 215, 2210 Eighth Street \\ Wright-Patterson AFB, OH 45433-7512, USA}

\cortext[cor1]{Corresponding author. Tel.:\ +1\ 610\ 758\ 4519.}

\begin{abstract}
The unsteady aerodynamic thrust and aeroelastic response of a two-dimensional membrane airfoil under prescribed harmonic motion are investigated computationally with a high-order Navier-Stokes solver coupled to a nonlinear membrane structural model.  The effects of membrane prestress and elasticity are examined parametrically for selected plunge and pitch-plunge motions at a chord-based Reynolds number of 2500.  The importance of inertial membrane loads resulting from the prescribed flapping is also assessed for pure plunging motions.  This study compares the period-averaged aerodynamic loads of flexible versus rigid membrane airfoils and highlights the vortex structures and salient fluid-membrane interactions that enable more efficient flapping thrust production in low Reynolds number flows. 
\end{abstract}

\begin{keyword}
flapping propulsion \sep membrane wing \sep micro air vehicle \sep vortex dynamics \sep Navier-Stokes simulations

\end{keyword}

\end{frontmatter}


\section{Introduction}
\input{Intro}

\section*{Nomenclature}
\input{nomen}

\section{Computational approach}
\input{Comp}

\section{Results}
    \subsection{Inertial Loading due to Rigid Body Motion}
\input{InertialLoads}
    \subsection{Thrust Enhancement from Membrane Elasticity and Prestress}
\input{ThrustIntro}
	\subsubsection{Plunging Motion}
\input{PlungeRes}
	\subsubsection{Pitch-Plunge Flapping Motion}
\input{PPres}

\section{Conclusions}
\input{Conc}

\section{Acknowledgments}
\input{Acknowledgment}

\renewcommand{\theequation}{A-\arabic{equation}}	
 
 \setcounter{equation}{0}  		
 
 \section*{Appendix: Estimation of Resonant Frequencies for a Fluid-Loaded Membrane}
\input{Appendix}

\bibliography{JFS_thrustaug}
\bibliographystyle{model2-names}

\end{document}

%% file: Intro.tex
Designers of micro air vehicles (MAVs) are currently looking to biological flight to address the technical challenges of low Reynolds number fliers.  These challenges include the aerodynamics of transitional flow, flight vehicle resilience to gusty environments, lightweight design, and propulsion.  Biological observations~\citep{paper:song:2008b}, experimental measurements~\citep{paper:mueller:2010}, and aeroelastic computations~\citep{paper:gordnier:2009b,paper:molki:2010} indicate that deformable membrane-airfoils may offer an engineering solution to these design issues, and that the existence of membrane flexibility may enable and enhance the thrust production of flapping wings in low Reynolds number flows.

In order to better understand the physics of a lifting membrane airfoil in a fluid flow, recent research has focused on the correlation between high-order aeroelastic simulations and experiment for the canonical configuration of~\citet{paper:rojratsirikul:2009,paper:rojratsirikul:2010b}, also considered in the present work, where a thin latex sheet is stretched tautly between two small, teardrop-shaped mounts; \citet{paper:arbos:2013} have extended experimental consideration to the effects of other leading- and trailing-edge mount shapes on the aeromechanics of such membrane airfoils at fixed angles of attack.  Previous computational works by \citet{paper:gordnier:2008} and \citet{paper:gordnier:2009a} have investigated membrane airfoil aeroelasticity over a range of steady angles of attack, membrane elasticity and prestress, and Reynolds numbers of up to 48\,500 where
laminar, transitional, and turbulent flows may be present simultaneously.  Their implicit large-eddy simulation (LES) approach~\citep{paper:visbal:2003} exploited a well-validated, robust, sixth-order Navier-Stokes solver to compute the three-dimensional flow field, where the aerodynamic forces are coupled to a finite element membrane model that captures geometric nonlinearity.  \citet{paper:jaworski:2012} applied the strong form of this membrane model to the high-order, two-dimensional aeroelastic solver by \citet{paper:gordnier:2009b} and extended the solver to include the rigid body motion of the membrane airfoil mounts.  This enhanced aeroelastic solver was then used to evaluate the aerodynamic performance of a two-dimensional membrane airfoil over a range of plunge amplitudes, prescribed flapping frequencies, and Reynolds numbers from 2500 to 10\,000.  

The present work employs the aeroelastic solver of \citet{paper:jaworski:2012}, fixing the Reynolds number at 2500 to maintain two-dimensional flow and to focus on the structural influences on the fluid-structure behavior.  These structural effects are namely the inertial loading due to prescribed rigid body motion and the elastic contributions from membrane elasticity versus prestress.  The inertial loading effects on thrust and propulsive efficiency are evaluated for pure plunging motions across a range of amplitudes and reduced frequencies.  The exploration of the elastic parameter space for pitching and pitch-plunge motions is motivated by the limited range of membrane prestress values examined in prior work that suggested important trade-offs between the prestress, required flapping power, and potential improvements to the period-averaged thrust and propulsive efficiency.  In the spirit of the exploratory study by~\citet{paper:streitlien:1998}, a `cross' in parameter space for the elastic modulus and prestress is examined to determine their importance on the vortex dynamics and aerodynamic performance, particularly the period-averaged thrust and propulsive efficiency, with the aim to inform the aeroelastic tailoring of more efficient flapping-membrane-airfoil fliers.

%% file: nomen.tex
\begin{tabbing}
$A$			\qquad \qquad	\=	cross-sectional area of membrane   \\
$c$					\>	membrane chord length   \\
$C_L$					\>	section lift normalized by $\rho_{\infty} V_{\infty}^2 c/2$   \\
$C_M$					\>	section moment about the pitch axis, positive nose-up, normalized by $\rho_{\infty} V_{\infty}^2 c^2/2$   \\
$C_p$					\>	pressure normalized by $\rho_{\infty} V_{\infty}^2 /2$  \\
$C_{\rm Power}$			\>	aerodynamic power normalized by  $\rho_{\infty} V_{\infty}^3 c/2$  \\
$C_T$					\>	section thrust normalized by  $\rho_{\infty} V_{\infty}^2 c/2$  \\
$\mathbf{d}$				\>	membrane deformation vector normalized by $c$, $\mathbf{d} = \lfloor u\mbox{,}w\rfloor^{\rm T}$   \\
$e_{xx}$				\>	membrane strain   \\
$E$					\>	elastic modulus normalized by $\rho_{\infty} V_{\infty}^2$   \\
$h^*$					\>	plunge amplitude normalized by $c$   \\
$h_{\rm s}$				\>	membrane thickness normalized by $c$   \\
$k$						\>	reduced frequency, $k=\omega c/(2 V_{\infty})$   \\
${\rm M}$						\>	Mach number   \\
${\rm Re}$				\>	Reynolds number based on membrane chord length and freestream velocity  \\
$^0\!S$					\>	membrane prestress normalized by $\rho_{\infty} V_{\infty}^2$   \\
$t$						\>	time   \\
$u$						\>	in-plane membrane deformation normalized by $c$   \\
$V_{\infty}$				\>	freestream fluid velocity   \\
$w$						\>	out-of-plane membrane deformation normalized by $c$   \\
$\mathbf{x}$				\>  position vector in local coordinate system normalized by $c$, $\mathbf{x}=\lfloor x\mbox{,}z\rfloor^{\rm T}$   \\
$\mathbf{x}_{\rm O}$			\>	normalized local position vector to pitch axis, $\mathbf{x}_{\rm O}=\lfloor x_0 \mbox{,}z_0 \rfloor^{\rm T}$   \\
$\mathbf{X}$				\>	position vector in global coordinate system normalized by $c$, $\mathbf{X} = \lfloor {\rm X}\mbox{,}{\rm Z}\rfloor^{\rm T}$   \\
$\mathbf{X}_{\rm R}$		\>	normalized global position vector to pitch axis of local coordinate system, \\
					\>	$\mathbf{X}_{\rm R}=\lfloor {\rm X_R}\mbox{,}{\rm Z_R}\rfloor^{\rm T}$   \\
$\alpha$				\>	geometric angle of attack	 \\
$\eta_{\rm p}$				\>	propulsive efficiency	 \\
$\rho_{\infty}$				\>	freestream fluid density 		\\
$\rho_{\rm s}$				\>	membrane density normalized by $\rho_{\infty}$	 \\
$\tau$					\>	nondimensional time, $\tau=V_{\infty} t/c$	 \\
$\bar{\tau}$				\>	nondimensional time scaled by the period of prescribed motion, $\bar{\tau}=\tau k/\pi$	 \\
$\mathbf{\Phi}$				\>	external loading on membrane normalized by $\rho_{\infty} V_{\infty}^2$, $\mathbf{\Phi} = \lfloor \Phi_x\mbox{,}\Phi_z \rfloor^{\rm T}$	 \\
$\omega$				\>	radian frequency	 \\
$\bar{(\,)}$				\>	period-averaged value	 \\
$(\,)'$					\>	spatial derivative; $\partial (\,)/\partial x$	 \\
$\dot{(\,)}$				\>	$\partial (\,)/\partial \tau$	 \\
$(\,)_{\infty}$				\>	freestream condition	 \\
\end{tabbing}

%% file: Comp.tex
A detailed account of the governing equations and computational methodologies employed in the present aeroelastic investigation have been given by~\citet{paper:jaworski:2012} and therefore only a brief description is given here. %

The flow about a dynamically deforming membrane is computed using the two-dimensional, compressible Navier-Stokes equations written in strong conservative form, incorporating a general time-dependent curvilinear coordinate transformation to allow for a deforming mesh. A compact finite-difference approach is employed to discretize the flow equations. An eighth-order filtering technique~\citep{paper:gaitonde:1997,paper:visbal:1999} is applied to the conserved variables to eliminate the long-term growth of spurious high-frequency modes associated with the compact discretization. The implicit, approximate-factorization algorithm of~\cite{paper:beam:1978} marches the flow equations forward in time, and this scheme is augmented by Newton-like subiterations to achieve second-order temporal and sixth-order spatial accuracy. 

The structural deformations of the elastic surface are assumed to be governed by a set of nonlinear membrane equations \citep{paper:carrier:1945,paper:carrier:1949} that allow for strain contributions from both in- and out-of-plane displacements $u$ and $w$, respectively. 
\begin{figure}
    \centering
        \includegraphics[width=0.60\textwidth]{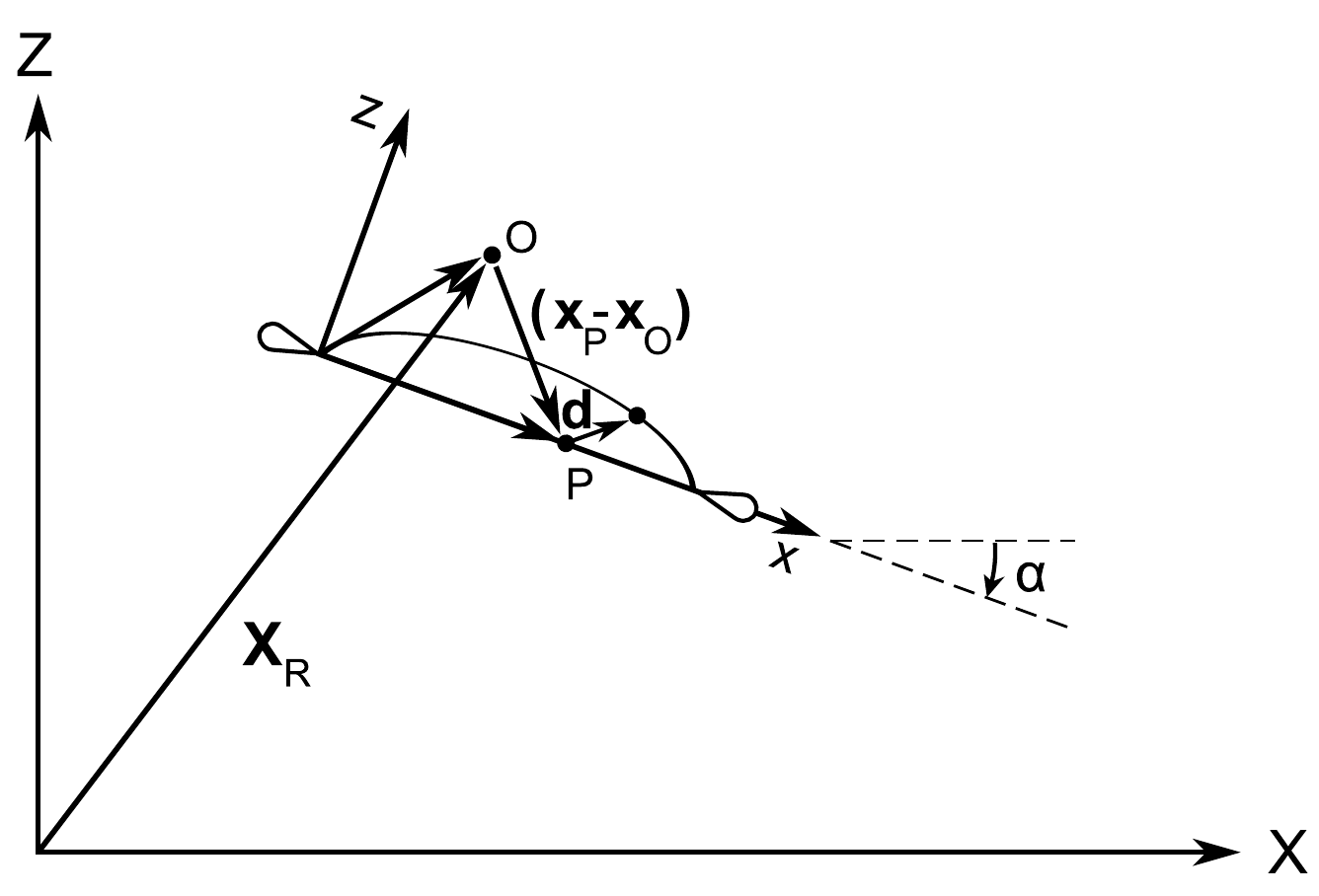}
    \caption[]{Global and local coordinate definitions for rigid body motion of the membrane airfoil. A uniform fluid flow with speed $V_{\infty}$ passes from left to right. Airfoil dimensions not drawn to scale.}
    \label{fig:RBMschematic}
\end{figure}
The membrane strain is defined geometrically by
\begin{equation}
e_{xx} = \left[ (1+u')^2 + {w'}^2 \right]^{1/2} - 1,
\end{equation}
and the constitutive relation between the membrane strain and tension is assumed to be linear elastic with a constant prestress in the undeformed state.
\begin{equation}
T = (^0\!S + E \, e_{xx}) A.
\end{equation}
The tension acts on area $A=h_{\rm s}\cdot 1$, the membrane thickness multiplied by a unit depth into the page. The resulting equations of motion are
\begin{align}
\rho_{\rm s} & h_{\rm s} \left[ \ddot{u} \, \underline{+ 2 \alpha \dot{w} - {\dot{\alpha}}^2 (u+x-x_0) +\ddot{\alpha} (w-z_0) } \right]
\nonumber \\
 &- \left\{ T \,(1+u') \left[ (1+u')^2 + (w')^2 \right]^{-1/2} \right\}' = \underline{- \rho_{\rm s} h_{\rm s} (\ddot{{\rm
X}}_{\rm R} \cos \alpha - \ddot{{\rm Z}}_{\rm R} \sin \alpha) } + \Phi_{x} \label{eq:Ueqn}\\
\rho_{\rm s} & h_{\rm s} \left[ \ddot{w} \, \underline{- 2 \alpha \dot{u} - \ddot{\alpha} (u+x-x_0) - {\dot{\alpha}}^2 (w-z_0) }
\right] \nonumber \\
 &- \left\{T \, w' \left[ (1+u')^2 + (w')^2 \right]^{-1/2} \right\}' = \underline{- \rho_{\rm s} h_{\rm s} (\ddot{{\rm X}}_{\rm R}
\sin \alpha + \ddot{{\rm Z}}_{\rm R} \cos \alpha) } + \Phi_{z} \label{eq:Weqn}
\end{align}
for a membrane airfoil whose rigid leading- and trailing-edge mounts undergo the rigid body motions shown in Fig.~\ref{fig:RBMschematic}. The prescribed time-dependent airfoil motions of angle of attack $\alpha$ and translational motion $\mathbf{X}_{\rm R}$ give rise to the underscored inertial load terms in Eqs.~(\ref{eq:Ueqn}) and (\ref{eq:Weqn}), which appear as body forces in the membrane-local frame. When considering pressure forces only, the external fluid loads on the membrane become $\Phi_{x} = -w' \Delta p$ and $\Phi_{z} = (1+u') \Delta p$ for a pressure difference $\Delta p = p^- - p^+$ across the membrane. 

The structural equations (\ref{eq:Ueqn}) and (\ref{eq:Weqn}) are solved in the membrane-local frame using standard finite-difference procedures. The temporal derivatives are computed using a second-order accurate three-point backward stencil, and the spatial derivatives are represented by second-order accurate central differences. Equations (\ref{eq:Ueqn}) and (\ref{eq:Weqn}) are linearized by evaluating the nonlinear structural terms in braces at the previous subiteration.  The external forces due to aerodynamic loads are assumed to be known from the previous subiteration.  Any degradation in the solution accuracy due to the linearization of Eqs.~(\ref{eq:Ueqn})
and (\ref{eq:Weqn}) and the lagging of the aerodynamic forces is eliminated via subiterations, which are also used in the solution of the aerodynamic equations.  Validation studies for both the structural and aeroelastic solvers have been carried out by \citet{paper:jaworski:2012}.

The dynamic two-way fluid-membrane coupling arises from the time-dependent fluid loads acting on the membrane surface and by the instantaneous structural displacements deforming the aerodynamic mesh. In the present work, only the pressure forces are passed to the membrane solver, and the deformation of the aerodynamic mesh due to membrane motion is
restricted to out-of-plane motions only. The simple algebraic method of \citet{paper:melville:2000} deforms the aerodynamic mesh to accommodate the new membrane surface. For the present computations, the spacings in the structural and aerodynamic solvers along the membrane surface are equal and no interpolation is required.  The temporal synchronization of the aerodynamic and membrane equations is achieved by the global subiteration procedure described previously, which eliminates temporal lag and the errors introduced by factorization and linearization of the governing equations. The resulting coupled procedure retains second-order temporal accuracy.

The computed membrane airfoil geometry is based on the experimental model of~\citet{paper:rojratsirikul:2009,paper:rojratsirikul:2010b}, shown schematically in Fig.~\ref{fig:RBMschematic}.  An O-type computational mesh represents the experimental geometry with 409 points round the membrane airfoil and 151 points in the normal direction, including 50-60 points in the boundary layer region.  The leading- and trailing-edge mounts are each described by 103 points, and 101 points describe the upper and lower flexible membrane surfaces with a maximum spacing of $\Delta x = 0.0149$. The minimum normal spacing at the membrane surface is $\Delta z_{\rm min} = 0.0001$, and the grid stretches to the outer boundary located 100 chords from the airfoil.  Freestream conditions are specified along the inflow portion of the farfield boundary, while a simple extrapolation of all variables is used on the outflow portion of the boundary. At the solid surface, the no-slip condition is applied, requiring that the fluid velocity at the airfoil surface match the surface velocity. In addition, the adiabatic wall condition and zero normal pressure gradient condition are specified. Unless otherwise noted, each flexible membrane computation assumes a mass ratio $\rho_{\rm s}h_{\rm s}=1.2065$, elasticity parameter $Eh_{\rm s}=306.98$, and prestress parameter $^0\!S h_{\rm s}=15.349$ to be consistent with previous computations~\citep{paper:attar:2011,paper:jaworski:2012}.  The membrane structural parameters are nondimensionalized in the aeroelastic solver by the fluid parameters, as detailed in the nomenclature.  The flow is essentially incompressible with a prescribed freestream Mach number of ${\rm M}_{\infty}=0.05$, and the Reynolds number is fixed at 2500.

%% file: InertialLoads.tex
The transformation of the membrane equation to a reference frame that undergoes prescribed translation and rotation produces inertial loads that, in addition to the fluid loading, affect the membrane deformations over the course of the flapping cycle (cf.\ Eqs.~(\ref{eq:Ueqn})--(\ref{eq:Weqn})).  The importance of these inertial loads on the period-averaged propulsive metrics of thrust, the power required to maintain the flapping motions, and the propulsive efficiency is evaluated here for prescribed pure plunge motions.  Each plunging computation assumes the prescribed motion ${\rm Z}_{\rm R}= -h^* \sin 2k \tau$ at zero geometric angle of attack within the following parameter space: $h^* = 0.25, 0.5$ and $k = 0.25, 0.5, 1.0$; the time-dependent effective angle of attack for these plunging motions is $2 k h^* \cos 2 k \tau$.  The plunge motions are chosen to augment the recent computations by \citet{paper:jaworski:2012}, which examine the parameter space between the static computations of \citet{paper:gordnier:2009b} and the dynamic computations of \citet{paper:attar:2011}.  The propulsive metrics are averaged over five simulated periods of prescribed motion and are compared to complementary flat, rigid membrane computations as well as thin airfoil theory predictions adapted from \citet{tech:theodorsen:1935} and \citet{tech:garrick:1936}, which have been recently collected and extended by \citet{paper:jaworski:2012a}.

\begin{figure}
    \begin{center}
        \includegraphics [width=0.60\textwidth]{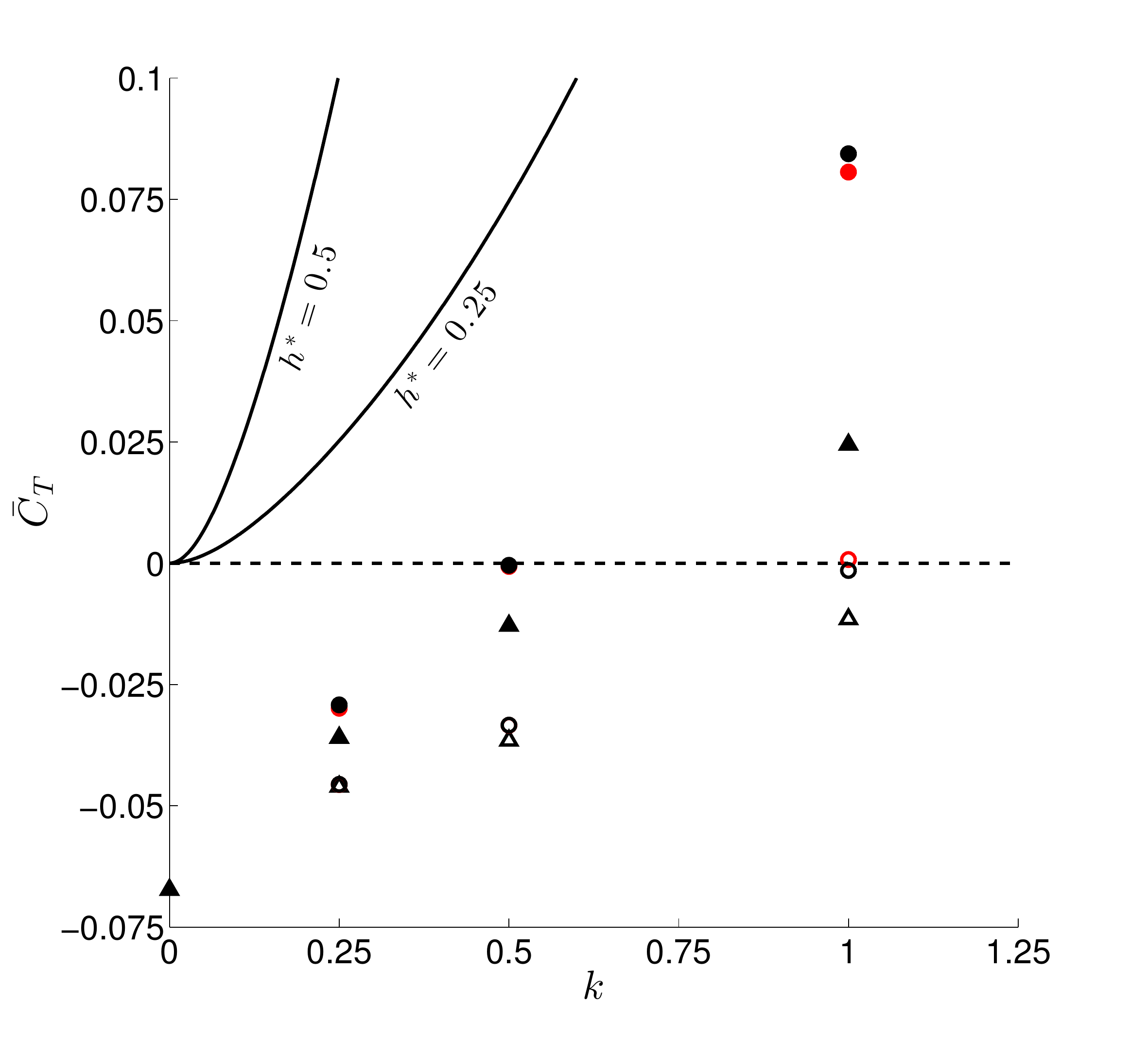} 
    \caption{Period-averaged thrust coefficient for plunging rigid and flexible membrane airfoils.  Open/filled symbols, $h^*=0.25, 0.5$; $\triangle$, rigid; $\circ$, flexible; $\textcolor{red}{\circ}$, flexible without rigid-body inertial loads; --, thin airfoil theory.   }
    \label{fig:CDmeanPlunge}
    \end{center}
\end{figure}

Figure~\ref{fig:CDmeanPlunge} compares the period-averaged thrust coefficients for a plunging airfoil, which always
produces an average propulsive force in classical thin airfoil theory.  However, in the present computations a positive net thrust is
achieved only under certain parameter combinations.  Drag is observed in the simulated plunge cases due in part to
viscous tractions on the airfoil surface, which are absent in thin airfoil theory because of its assumption of inviscid
potential flow. All computations at low frequency $k=0.25$ produce a net drag, which is greatest for the rigid airfoil
undergoing small amplitude motions.  Larger plunge motions and plunge frequencies contribute to
greater thrust (or lesser drag) in agreement with the trends of linear theory.  Also, membrane flexibility leads
consistently to a drag reduction or thrust augmentation when compared to rigid airfoil performance.  Comparing the flexible membrane results, the mean thrust is shown to be virtually unaffected by the removal of rigid-body inertial loads over the range of considered plunge amplitudes and frequencies.  The difference between the flexible membrane mean thrust results with or without rigid-body inertial loads grows with increasing reduced frequency as expected by the dependence of these loads on the acceleration of the reference frame; however, this difference in $\bar{C}_T$ remains comparatively small over the range of amplitudes and frequencies considered.

\begin{figure}
    \begin{center}
    \includegraphics[width=0.59\textwidth]{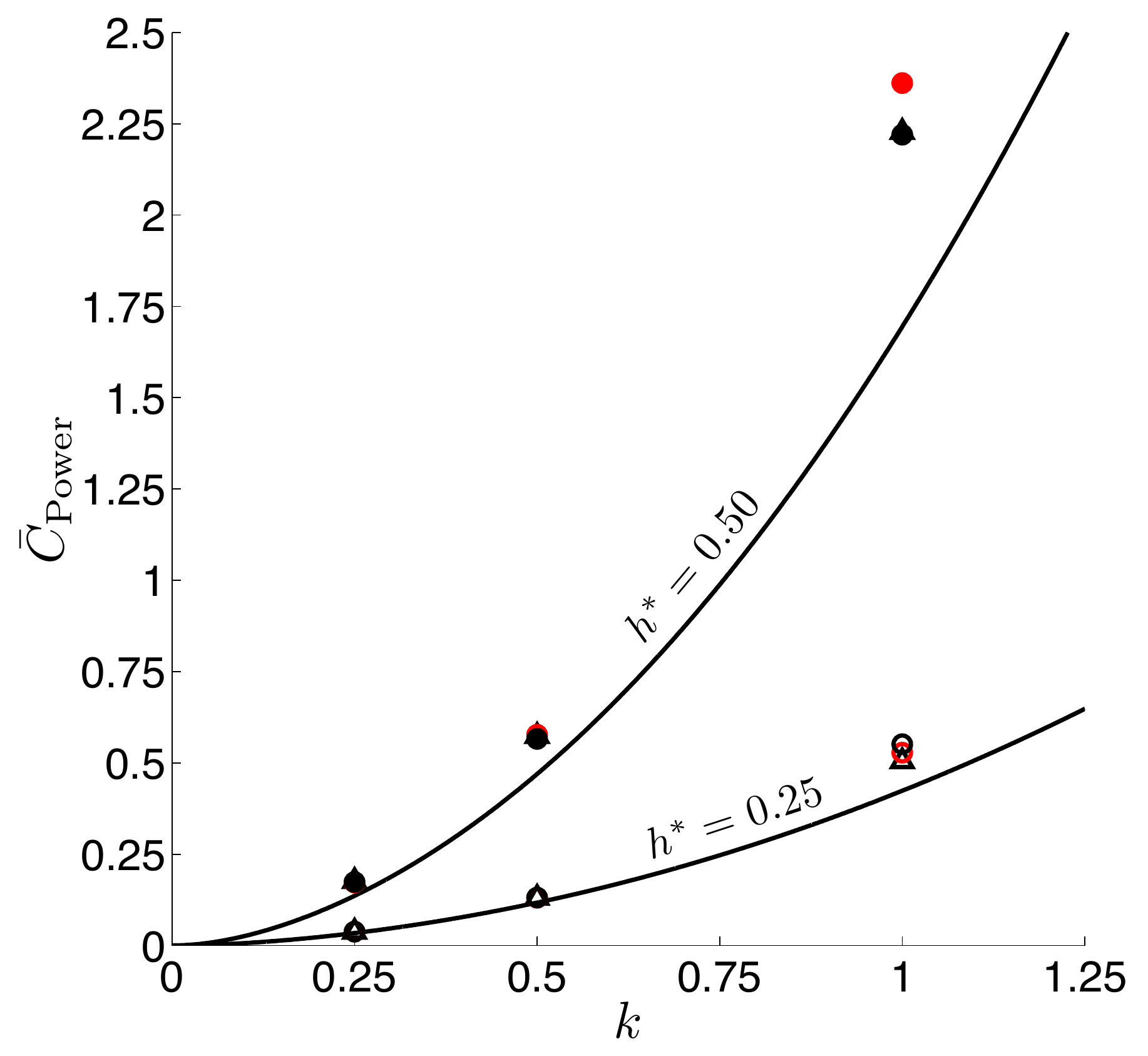}
    \caption{Period-averaged coefficient of required aerodynamic power versus reduced frequency for plunging rigid and flexible membrane airfoils. Open/filled symbols, $h^*=0.25, 0.5$; $\triangle$, rigid; $\circ$, flexible; $\textcolor{red}{\circ}$, flexible without rigid-body loads; --, thin airfoil theory.
    }
    \label{fig:CPower}
    \end{center}
\end{figure}

The period-averaged aerodynamic power necessary to sustain both pitching and plunging harmonic motions may be expressed as
\begin{equation}
\bar{C}_{\rm Power} = -\frac{k}{\pi} \int_0^{\pi/k} -\dot{h}^*(\tau) \, C_L (\tau)+ \dot{\alpha}(\tau) \, C_M(\tau) \, d\tau,
\label{eq:Cpower}
\end{equation}
where the power requirement for pure plunge motions depends entirely on the lift and prescribed motion. Note that Eq.~(\ref{eq:Cpower}) does not include membrane inertial loads or strain energy contributions, and therefore $\bar{C}_{\rm Power}$ is strictly a measure of the power necessary to act against the aerodynamic loads.  Figure~\ref{fig:CPower} compares this computed aerodynamic power requirement against predictions from thin airfoil theory. As in the mean thrust results, there is virtually no difference between the required power to act against the aerodynamic loads for the flexible membrane cases with or without the inertial loads as affected by the aeroelastic interaction (recalling that membrane inertial forces are not included in the power calculation), with the exception of $k=1.0$ where the difference is distinct but still small (less than $6.5$\%).  Close numerical agreement between the rigid and flexible membrane power requirements indicates that membrane flexibility plays a minor role in the required power to sustain pure plunge flapping for the range of  amplitudes and frequencies considered.  Good agreement is also noted between the simulation results and the predictions from linear theory.  The simulations consistently require a greater amount of power than theory suggests, which indicates that the predictions from linear theory may be thought of as a lower bound for the power coefficient.

\begin{figure}
    \begin{center}
    \includegraphics [width=0.59\textwidth] {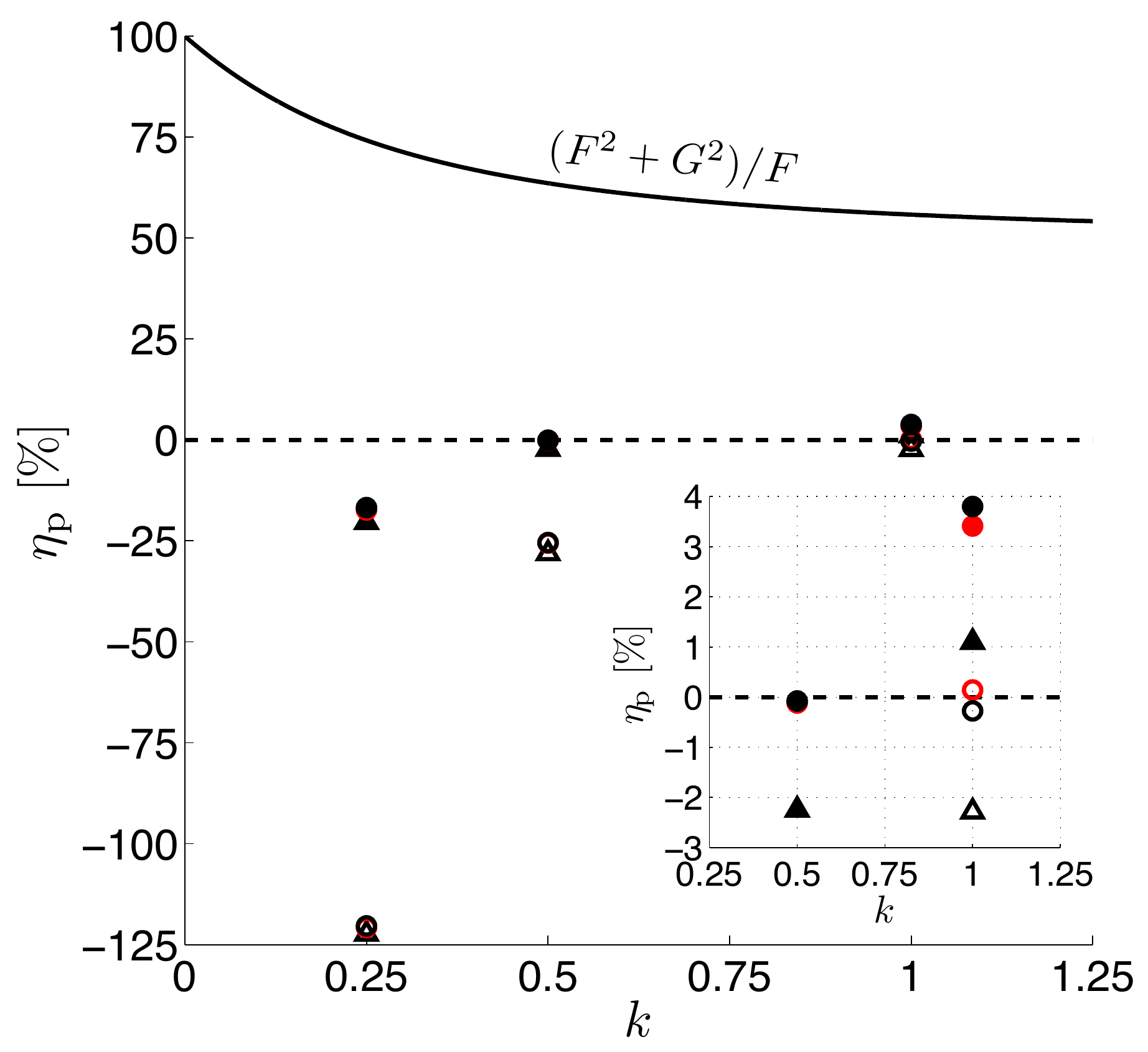}
    \caption{Propulsive efficiency for plunging rigid and flexible membrane airfoils.  Legend as in Figure~\ref{fig:CPower}. The thin airfoil result is expressed in terms of the real and imaginary parts of Theodorsen's function, $C(k) = F + {\rm i}G$ \citep{tech:theodorsen:1935}.}
    \label{fig:etap}
    \end{center}
\end{figure}

Figure~\ref{fig:etap} compares the theoretical and computed propulsive efficiencies, defined here as $\eta_{\rm
p}=\bar{C}_T/\bar{C}_{\rm Power}$.  As noted by \citet{paper:jaworski:2012}, the variations in computed propulsive efficiency are largely attributed to the
thrust coefficient because the differences in $\bar{C}_{\rm Power}$ are comparatively much smaller.  Inviscid thin
airfoil theory predicts perfect propulsive efficiency at infinitely slow frequency and a minimum efficiency of 50\% for
very rapid flapping. As stated previously, a membrane airfoil in a real flow must
overcome viscous fluid forces to generate a net thrust, and therefore greater propulsive efficiency is achieved as
flapping amplitude and frequency increase. The maximum propulsive efficiency achieved over the present range of amplitudes and frequencies is
low ($\eta_{\rm p} \approx 4\%$).  As anticipated by the small variations in mean thrust and power coefficients for membrane airfoil simulations with or without the rigid-body inertial loads, the change in propulsive efficiency due to these loads is also small, with a maximum deviation of $\Delta \eta_{\rm p} = 0.42\%$ for $h^*=0.25, k=1.0$.

In summary, the inertial loads resulting from rigid-body motions have a minor effect on the period-averaged propulsive metrics for the range of plunging motions considered.  In fact, the simulation results with or without the inertial loads are virtually indistinguishable from each other with the exception of the high frequency cases at $k=1.0$ where only modest differences are observed. Clearly, the fluid and elastic-structural forces dominate the dynamic body forces in Eqs.~(\ref{eq:Ueqn}) and (\ref{eq:Weqn}) for the range of plunging cases considered.

%% file: ThrustIntro.tex
The influence of membrane flexibility, as affected by the modulus of elasticity and prestress, on the period-averaged propulsive characteristics of membrane airfoils is examined for selected plunge and pitch-plunge motions.  The flapping motions are simulated within the following `cross' in parameter space, where either $E h_{\rm s}=306.98$ or $^0\!S h_{\rm s} = 15.349$ is held fixed while the other is varied: $E h_{\rm s} = 50, 100, 200, 306.98, \infty$ (rigid); $^0\!S h_{\rm s} = 2, 5, 10, 15.349, \infty$ (rigid). In the limit of the rigid membrane, the reader is directed to the discussion by~\citet{paper:jaworski:2012} regarding the effects of airfoil geometry on thrust in comparison with the numerical study of~\citet{paper:ashraf:2011} and the experimental work of~\citet{paper:anderson:1998}.

%% file: PlungeRes.tex
The plunging motion parameters are now set to $h^* = 0.5$ and $k=1.0$.  For this motion, the increase in period-averaged thrust has been shown by \citet{paper:jaworski:2012} to arise from the low pressure regions due to both the attached and shed leading-edge vortices.  The dominant thrust contribution comes from the leading edge vortex inducing and acting upon the membrane camber.  The shed vortex in proximity to the 70\%-chord region mitigates the pressure drag on the airfoil and can also provide a suction force to contribute to the net thrust~\citep{paper:jaworski:2012}.  

\begin{figure}
  \begin{center}
   \mbox{
    \subfigure[]{\scalebox{0.31}{ \rotatebox{0}{\includegraphics{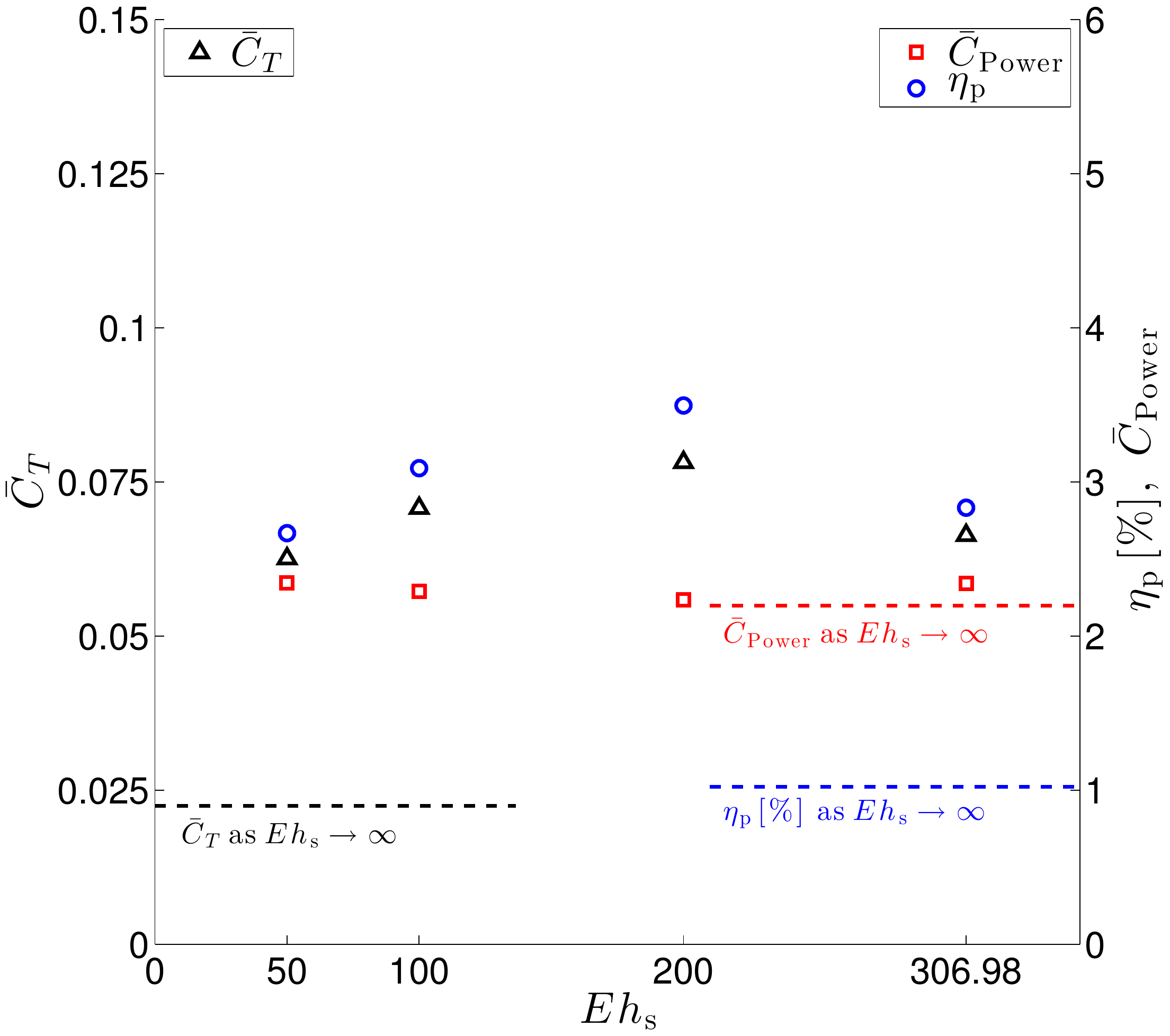}}} \label{fig:PlungeEH}
} 
    \subfigure[]{\scalebox{0.31}{ \rotatebox{0}{\includegraphics{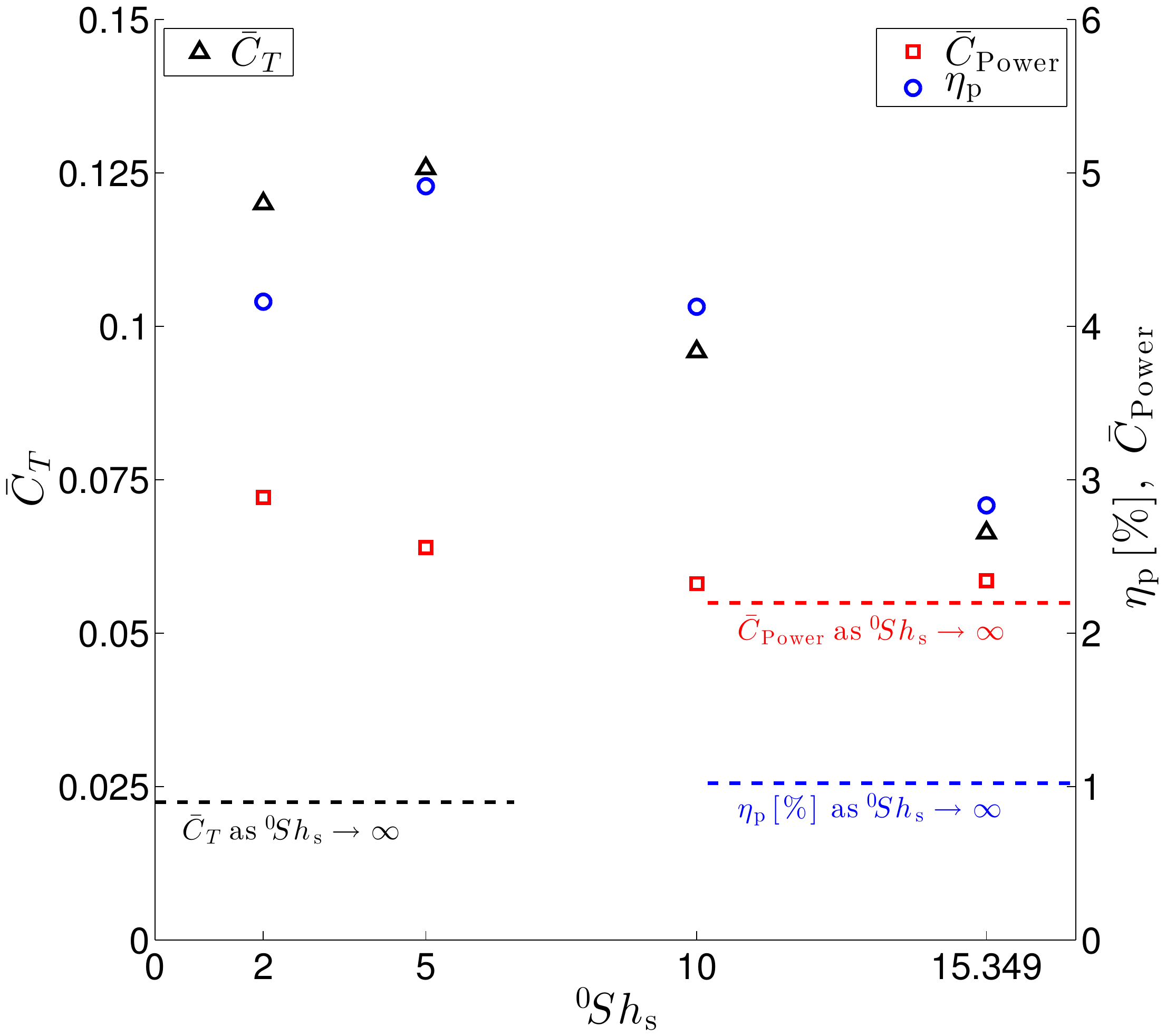}}} \label{fig:PlungeSH}
}
    }
  \end{center}
\caption[]{Plunging motion propulsion metrics versus membrane flexibility parameters: (a) elasticity; (b) prestress.  $h^*=0.5, k=1.0$.}
\label{fig:PlungeFlex}
\end{figure}

\begin{figure}
    \begin{center}
    \includegraphics [width=0.59\textwidth] {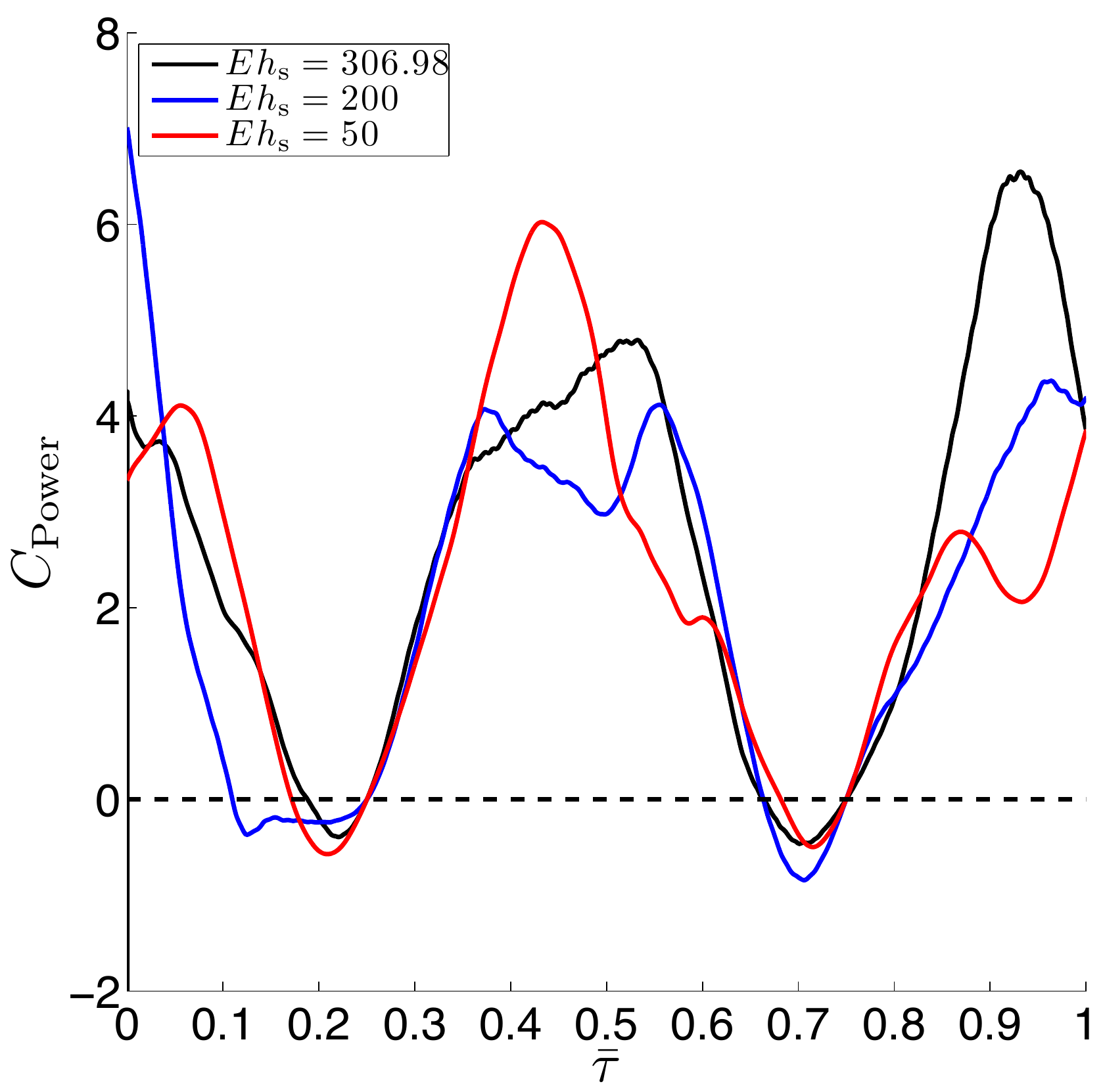}
    \caption{Instantaneous power coefficient for plunging flexible membrane airfoil. $h^*=0.5, k=1.0, ^0\!Sh_{\rm s} = 15.349$.}
    \label{fig:CPower_instant}
    \end{center}
\end{figure}

\begin{figure}
    \begin{center}
    \includegraphics [width=1.0\textwidth] {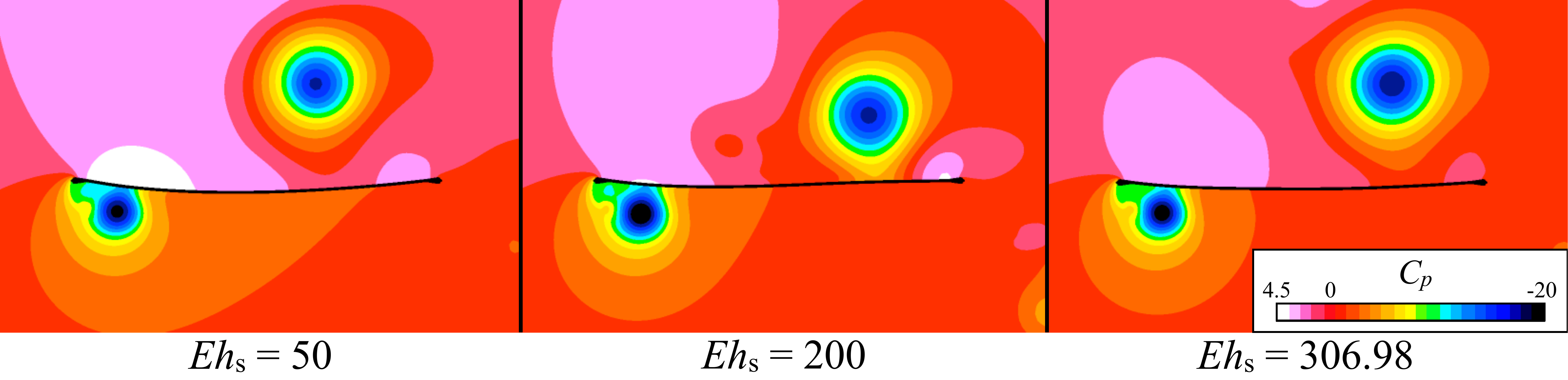}
    \caption{Pressure contours about a membrane airfoil at $\bar{\tau} = 0.425$ at various values of elastic modulus. $h^*=0.5, k=1.0, ^0\!Sh_{\rm s} = 15.349$.}
    \label{fig:CP_plungeEH}
    \end{center}
\end{figure}

Figure~\ref{fig:PlungeEH} shows that when the elastic modulus parameter is varied the maximum thrust and propulsive efficiency coincide with the minimum required aerodynamic power for $E h_{\rm s}=200$.  A representative period of the instantaneous power coefficient in Fig.~\ref{fig:CPower_instant} reveals that greatest reduction in required power comes from minimizing the unsteady lift every half-period where its value is most important, noting here that $C_{\rm Power} = 2 k h^* \, C_L(\bar{\tau}) \cos 2 \pi \bar{\tau}$.  Figure~\ref{fig:CP_plungeEH} illustrates the pressure field about the membrane airfoil for $Eh_{\rm s} = 50$,  $200$, and $306.98$ for $\bar{\tau} = 0.425$, at which point in the cycle the airfoil is moving upwards.  The pressure distributions are similar for all three cases along the bottom surface of the membrane, whose nearby flow field is dominated by the attached leading edge vortex.  Thus, the only significant differences in instantaneous lift are due to the pressure distribution on the upper surface influenced by the large vortex shed previously in the flapping cycle.

The case of $Eh_{\rm s}=200$ maintains this vortex closer to the membrane surface than for $Eh_{\rm s}=50$ or $306.98$, leading to a less negative integrated lift force over the membrane and a lower power requirement.  Also, it is noted that the increase in the elastic modulus diminishes the pressure differential near the leading edge mount, further contributing to the reduction of the lift magnitude and required aerodynamic power.  Therefore, the combined attached-leading- and shed-vortex mechanism to enhance the thrust of a plunging airfoil at $h^*=0.5$ and $k=1.0$ also reduces the instantaneous lift every half-period, leading to a reduced period-averaged power requirement and greater propulsive efficiency that is sensitive to the membrane elasticity.

\begin{figure}
    \begin{center}
    \includegraphics [width=0.59\textwidth] {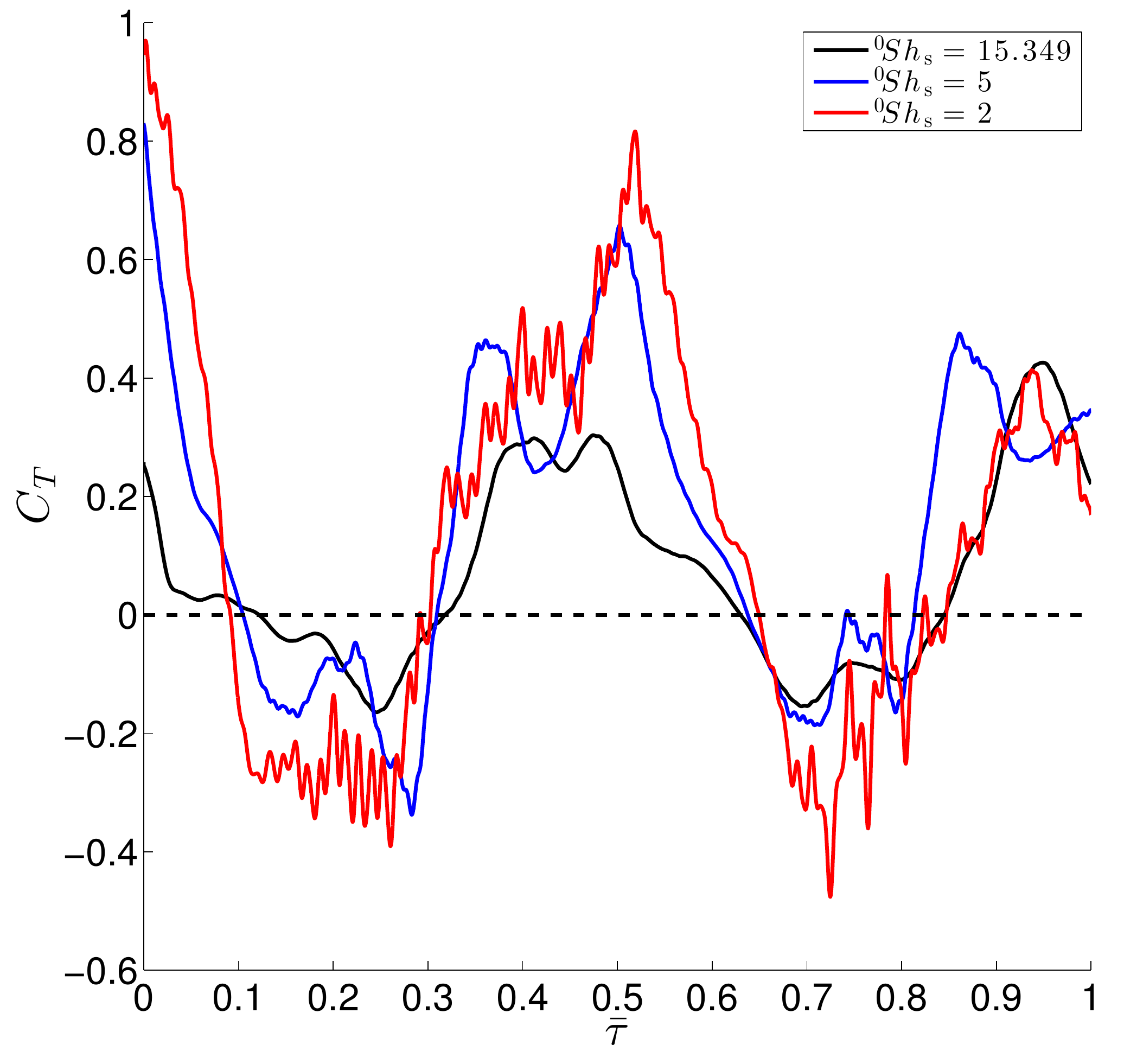}
    \caption{Instantaneous thrust coefficient for plunging flexible membrane airfoil. $h^*=0.5, k=1.0, Eh_{\rm s} = 306.98$.}
    \label{fig:CDrag_instant}
    \end{center}
\end{figure}

\begin{figure}
    \begin{center}
    \includegraphics [width=1.0\textwidth] {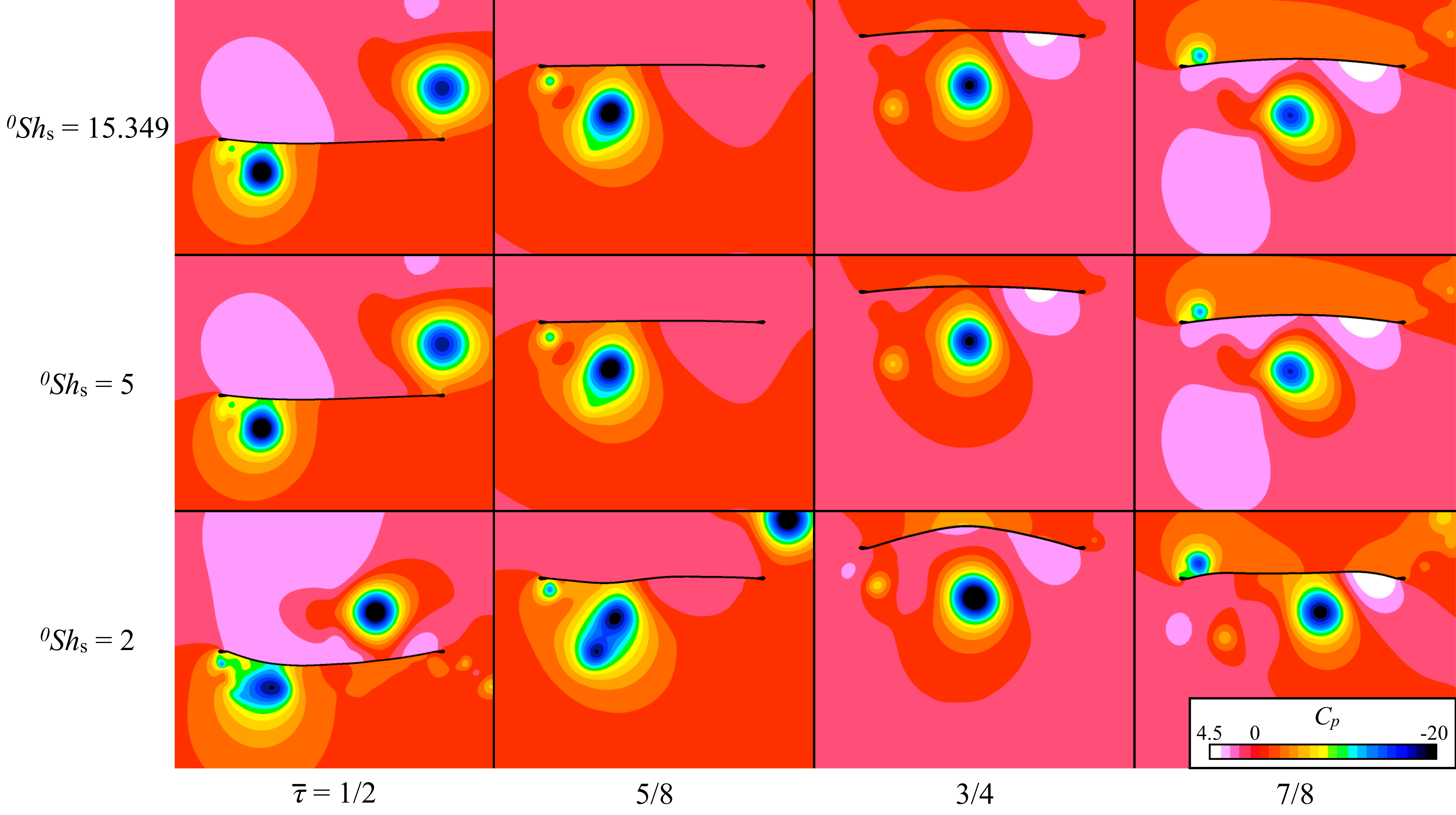}
    \caption{Pressure contours about a plunging membrane airfoil at various values of prestress. $h^*=0.5, k=1.0, Eh_{\rm s} = 306.98$.}
    \label{fig:CP_plungeSH}
    \end{center}
\end{figure}

When varying the membrane prestress, Fig.~\ref{fig:PlungeSH} demonstrates that the thrust and propulsive efficiency have a maximum for $^0\!S h_{\rm s} = 5$ and that the required power grows monotonically as prestress decreases.  The comparison of representative thrust histories in Fig.~\ref{fig:CDrag_instant} for three finite prestress cases shows how decreasing prestress leads to stronger thrust events near each half-period as well as the emergence of an additional high-frequency response and a diminished average thrust for $^0\!Sh_{\rm s}=2$.  At present it is not conclusive whether or not the high-frequency content of the thrust response is due to particular flow features, an excitation of certain membrane modes, or some other physical phenomenon. The excitation or forced flapping of elastic aerodynamic structures at or near their fluid-loaded resonances has been shown by recent studies~\citep{paper:michelin:2009, paper:kang:2011, paper:alben:2012, paper:leftwich:2012, paper:dewey:2013} to enhance thrust production. Analytical estimates of the fundamental frequency of the fluid-loaded membrane are carried out in the Appendix and range from $k=1.282$ to 4.137 for ${}^0\!S h_{\rm s} = 2$ to 15.349, respectively. Thus, the aeroelastic response of the membrane airfoil is likely affected by proximity to resonant forcing for $k=1.0$ with smaller values of nondimensional prestress. Accordingly, the pressure contours about the membrane airfoil in Fig.~\ref{fig:CP_plungeSH} indicate that more complicated membrane deformations occur over the cycle due to the greater influence of the fluid versus elastic forces as prestress is decreased.  Indeed, these irregular deformations can lead to a strong dependence of the fluid and structural dynamics on their time histories, in contrast to the well-behaved and periodic drag response for $^0\!Sh_{\rm s} = 15.349$ where the membrane deformation deviates only slightly from a parabolic shape over the course of the flapping cycle due to the leading edge vortex.  Interestingly, in addition to the maxima in propulsive force and efficiency for the selected plunge motion, the rigid limits of $Eh_{\rm s} \rightarrow \infty$ or $^0\!Sh_{\rm s} \rightarrow \infty$ are consistently found to be the lower bounds for all of the propulsion metrics in Fig.~\ref{fig:PlungeFlex}.

%% file: PPres.tex
The combined pitch and plunge motions for the membrane airfoil are based on the experiment by~\citet{paper:anderson:1998}  The airfoil motion is defined by ${\rm
Z}_{\rm R}= -h^* \sin 2k \tau$ and $\alpha = 30^{\circ} \sin (2k \tau+ 90^{\circ})$ about a zero mean angle of attack, where $h^*=-1.0$ and $k=0.71$. The pitch axis is one-third chord distance aft
of the airfoil leading edge. 

\begin{figure}
  \begin{center}
   \mbox{
    \subfigure[]{\scalebox{0.31}{ \rotatebox{0}{\includegraphics{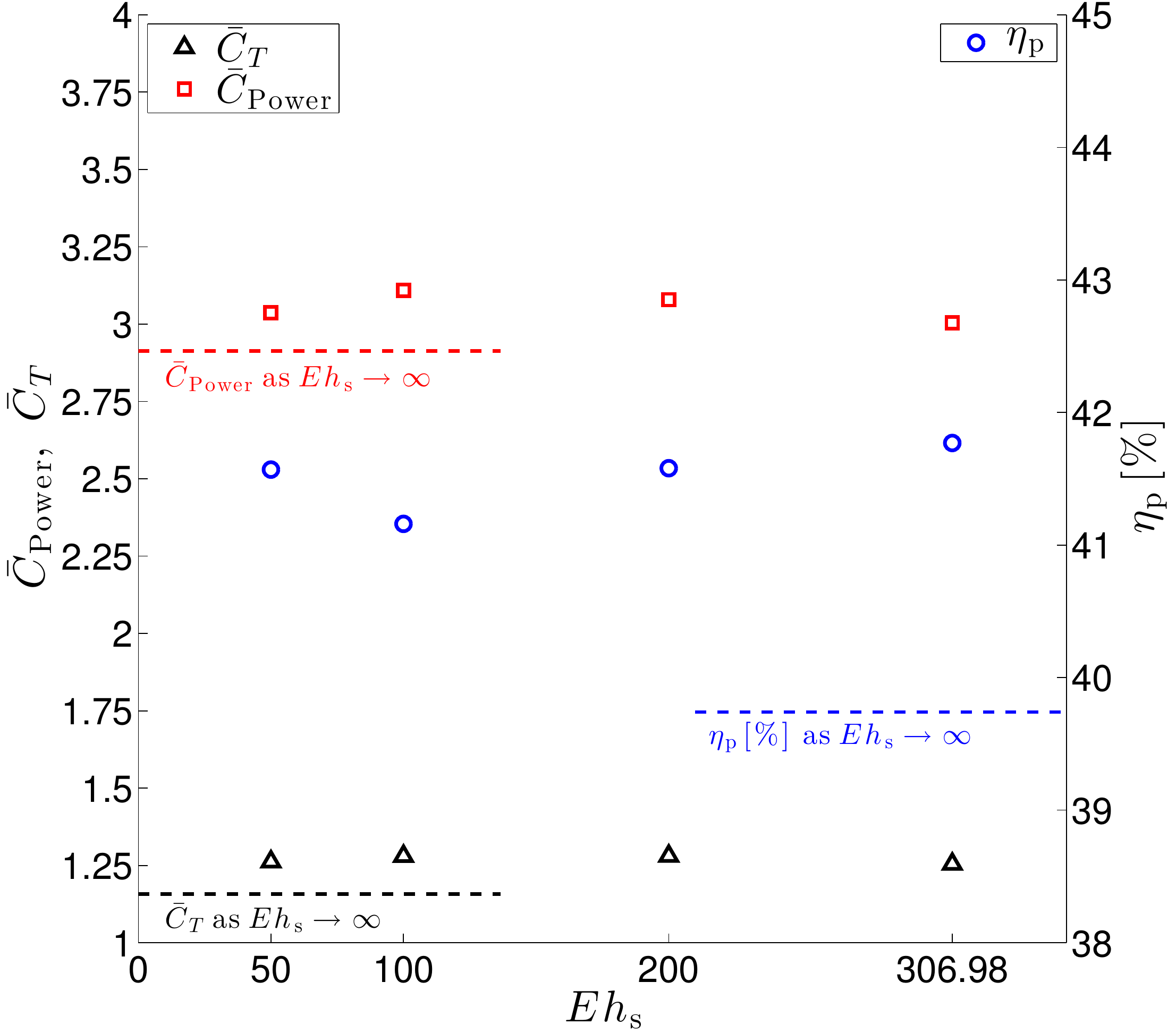}}} \label{fig:PP_EH}
} 
    \subfigure[]{\scalebox{0.31}{ \rotatebox{0}{\includegraphics{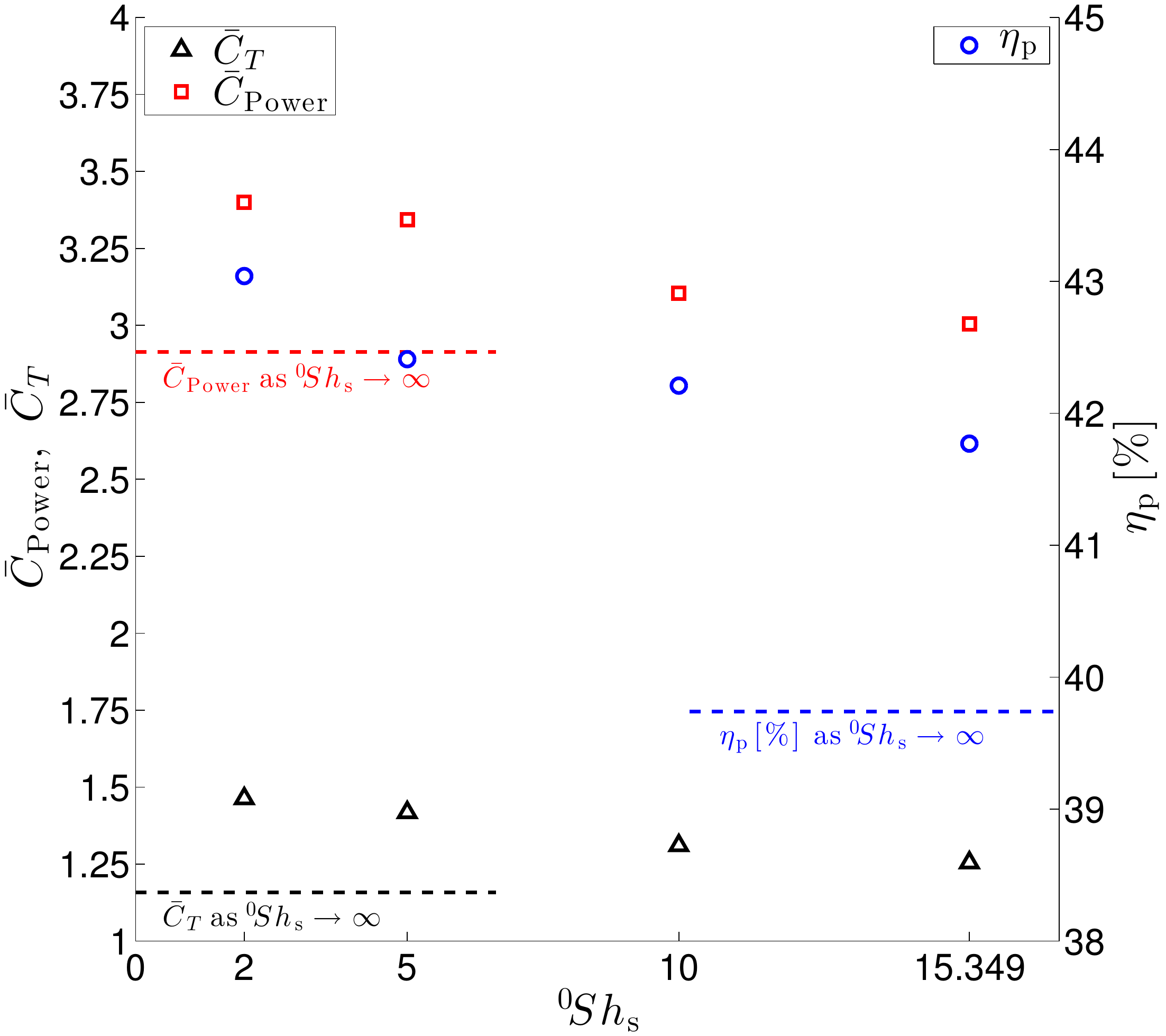}}} \label{fig:PP_SH}
}
    }
  \end{center}
\caption[]{Pitch-plunge flapping propulsion metrics versus membrane flexibility parameters: (a) elasticity; (b) prestress. ${\rm
Z}_{\rm R}= \sin 2k \tau$, $\alpha = 30^{\circ} \sin (2k \tau+ 90^{\circ})$, $k=0.71$.}
\label{fig:PPflex}
\end{figure}

As observed for the pure plunge motion, Fig.~\ref{fig:PPflex} shows that the rigid limit again furnishes a lower bound for the propulsive efficiency as well as for the coefficients of thrust and power.  Noting the expanded vertical axes in Fig.~\ref{fig:PP_EH}, the thrust coefficient and propulsive efficiency are relatively flat across the range of $E h_{\rm s}$.  However, the relative maximum propulsive efficiency at $Eh_{\rm s}=306.98$ and jump in $\eta_{\rm p}$ between $E h_{\rm s} = 306.98$ and $\infty$ (rigid) suggest that a global maximum in propulsive efficiency may lie between these values.  The relative minimum propulsive efficiency at $E h_{\rm s} = 100$ is due to a required power maximum, but the principal conclusion from Fig.~\ref{fig:PP_EH} is that the propulsion due to the prescribed pitch-plunge motion for $^0\!Sh_{\rm s} = 15.349$ is relatively insensitive to variations in the elastic modulus.

The variation in prestress for the pitch-plunge airfoil follows the same trends as for the plunging membrane airfoil; namely, the reduction in membrane prestress leads to the monotonic increase in the propulsive efficiency as well as the coefficients of thrust and required power.  As explained by \citet{paper:jaworski:2012}, the increase in required power is outpaced by the increased thrust due to the coupled interaction of the stronger leading-edge vortex inducing a greater membrane camber as prestress decreases.

%% file: Conc.tex
A high-order, two-dimensional aeroelastic solver is used to evaluate the period-averaged drag of a membrane airfoil under plunge and pitch-plunge flapping motions for Reynolds number $2500$.  Numerical simulations of pure plunge motions are carried out for $h^*=0.25, 0.5$ and $k=0.25, 0.5, 1.0$ for a flexible membrane either with or without the inertial loads that result from rigid-body translation of the airfoil.  These inertial loads are shown to have virtually no effect on the thrust results for the parameters considered.

The effects of varying the membrane elastic modulus and prestress are investigated by a `cross' in parameter space for $E h_{\rm s} = 50, 100, 200, \underline{306.98}, \infty$ and $^0\!S h_{\rm s} = 2, 5, 10, \underline{15.349}, \infty$, where the baseline case is identified by the underlined values.  When subjected to a particular pitch-plunge flapping motion, the variation in propulsive thrust, efficiency, and required aerodynamic power is small for the range of elastic moduli considered.  However, these propulsive metrics are found to increase monotonically as the membrane prestress decreases, noting that the mutually-reinforcing aeroelastic interaction between the attached leading-edge vortex and the membrane deformation leads to an improvement of the propulsive force that increases at a greater rate than the power coefficient as $^0\!S h_{\rm s}$ decreases.

Pure plunge motions also have a monotonic increase in required power for decreasing prestress, but at $^0\!S h_{\rm s} = 2$ the fluid forces begin to dominate the elastic restoring forces in the membrane, leading to complex membrane deformations and a high-frequency oscillation in and a high mean value of the thrust response.  Maxima for thrust and propulsive efficiency are obtained for $^0\!S h_{\rm s} = 5$ when varying membrane prestress and for $E h_{\rm s} = 200$ over the range of elastic moduli.  For this value of elastic modulus, the required power to sustain the flapping motions is minimized by the action of a shed leading-edge vortex reducing the magnitude of the unsteady lift every half-period.  This vortex is also responsible for the mitigation of pressure drag on the membrane and can interact favorably with the local membrane camber to enhance the propulsive force.  In both the plunge and pitch-plunge parametric variations of prestress and elastic modulus, the limit of a rigid membrane airfoil yields the lowest values of propulsive force and efficiency, supporting the potential of compliant wing surfaces to enhance the performance of flapping MAVs in low Reynolds number flows.

%% file: Acknowledgment.tex
This research was performed while the first author held a National Research Council Research Associateship Award at Wright-Patterson Air Force Base.  This work was sponsored in part by the Air Force Office of Scientific Research under a task monitored by Dr.\ D.\ Smith.  The first author also acknowledges support from the National Science Foundation under Award No.\ 0965248.  The insightful comments by the anonymous reviewers are gratefully acknowledged.

%% file: Appendix.tex
The nondimensional governing equation for a linear membrane aligned with a uniform potential flow of speed $V_{\infty}$ is
\begin{equation}
\rho_{\rm s} h_{\rm s} \ddot{w} - ^0\!\!S h_{\rm s} w'' = \frac{2}{\pi} \int_0^1 \frac{\dot{w} + w'}{x - \xi} + \left( \ddot{w} + \dot{w}' \right) \ln |x-\xi| \, d \xi.
\label{eq:fluidload}
\end{equation}
The integral term represents the non-circulatory fluid forces resulting from structural motion (see \citet{book:BAH}, p.~213). This potential flow assumption was determined by~\citet{paper:kornecki:1976} to furnish an adequate aerodynamic model for a flexible panel with fixed end conditions. Note that the Cauchy principal value should be taken when encountering improper integrals in Eq.~(\ref{eq:fluidload}). 

The resonant frequencies of the fluid-loaded membrane are determined by the procedure of \citet{paper:kornecki:1976}, whereby a separable solution form for the membrane vibration is assumed, 
\begin{equation}
w(x,t) = e^{i 2 k \tau}  \hat{w} (x),
\label{eq:eigsub}
\end{equation}
and the spatial function $\hat{w}(x)$ is expanded in terms of the \emph{in vacuo} structural modes,  $w_n(x) = \sin n \pi x $,  i.e.\
\begin{equation}
\hat{w}(x) = \sum_{n=1}^{\infty} C_n w_n(x).
\label{eq:what}
\end{equation}
Substitute equations (\ref{eq:eigsub}) and (\ref{eq:what}) into (\ref{eq:fluidload}) and apply Galerkin's method to obtain an infinite set of homogeneous algebraic equations in the constants $C_n$. This set of equations yields an eigenvalue problem for the nondimensional fluid-loaded natural frequency $k$,
\begin{equation}
\det || -4 k^2 (\rho_{\rm s} h_{\rm s} \delta_{mn} - D_{mn} ) - i 2 k B_{mn} + n^2 \pi^2 \, { }^0 \! S h_{\rm s} \delta_{mn} - A_{mn} || = 0, \quad \quad m,n = 1,2,3, \ldots,
\end{equation}
where $\delta_{mn}$ is the Kronecker delta,
\begin{align}
A_{mn} &= \int_0^1 I_0 (\hat{w}_m,x) w_n(x) \,  dx, \\
B_{mn} &= \int_0^1 I_1 (\hat{w}_m,x) w_n(x) \,  dx, \\
D_{mn} &= \int_0^1 I_2 (\hat{w}_m,x) w_n(x) \,  dx, 
\end{align}
and
\begin{align}
I_0(\hat{w}_n,x) &= \frac{4}{\pi}  \int_0^1  \frac{d \hat{w}_n/d\xi}{x-\xi} \, d\xi, \\
I_1(\hat{w}_n,x) &= \frac{4}{\pi}  \int_0^1  \frac{\hat{w}_n(\xi)}{x-\xi} + \frac{d \hat{w}_n}{d \xi} \ln |x-\xi| \, d\xi, \\
I_2(\hat{w}_n,x) &= \frac{4}{\pi}  \int_0^1  \hat{w}_n(\xi) \ln |x-\xi| \, d\xi.
\end{align}
\begin{table}
  \begin{center}
\def~{\hphantom{0}}
  \begin{tabular}{lcc}
  \hline \hline 
      ${ }^0 \! S h_{\rm s}$  & \multicolumn{2}{c}{$k$}  \\  \cline{2-3}
      						& \emph{In vacuo} & Fluid-loaded \\ \hline 
       15.349   				& 5.603 		& 4.137 \\
       10   					& 4.522 		& 3.303 \\
       ~5  					& 3.198 		& 2.261 \\
       ~2 					& 2.022 		& 1.282 \\ \hline \hline
  \end{tabular}
  \caption{Reduced natural frequencies for the first mode of a membrane airfoil with and without the effects of fluid loading. Ten structural modes were used to evaluate the fluid-loaded cases to the indicated precision. $\rho_{\rm s} h_{\rm s}  = 1.2065$. }
  \label{tab:natfreq}
  \end{center}
\end{table}
The natural frequencies of the membrane under fluid loading are computed and compared to the \emph{in vacuo} predictions in Table~\ref{tab:natfreq} for the range of nondimensional prestress considered in this work. It must be noted that these estimates are made without consideration of circulatory effects, which could be included using either the method of~\citet{paper:kornecki:1976} or~\citet{paper:huang:1995}, but these effects have typically a marginal influence on the fluid-structure interaction for the two-dimensional configuration considered here~\citep{paper:eloy:2007}. The comparison of resonant frequency predictions in Table~\ref{tab:natfreq} indicates that the presence of the background fluid flow reduces the membrane resonant frequency by 26-37\%, the magnitude of which depends on the prestress parameter. It is noteworthy that the natural frequencies are at least a factor of 1.282 or greater than the highest prescribed reduced frequency ($k=1.0$) in this study. Therefore, the fluid-membrane system may be in proximity to resonant forcing, where significant increases in aerodynamic thrust have been predicted~\citep{paper:michelin:2009, paper:kang:2011, paper:alben:2012} and observed~\citep{paper:alben:2012,paper:leftwich:2012,paper:dewey:2013} for flexible flapping systems. Recent research by~\citet{paper:moored:2014} suggests further that optimal thrust for a flapping system may be linked to the resonance between the fluid-loaded structure and its wake.